\documentclass{article}
\usepackage{spconf,amsmath,graphicx,multirow,array}
\usepackage{epstopdf}
\usepackage{amssymb}
\usepackage{caption}
\usepackage{algorithm} 
\usepackage{algorithmic} 
\usepackage{color}
\usepackage{hyperref}
\usepackage{mathrsfs} 
\usepackage{booktabs}
\usepackage{cite}
\definecolor{note}{RGB}{105, 105, 105} 
\usepackage{setspace}

\usepackage{subfigure}

\title{3S-TSE: Efficient Three-Stage Target Speaker Extraction for Real-Time and Low-Resource Applications}
\name{
\begin{tabular}{@{}c@{}}
Shulin He\textsuperscript{1}, Jinjiang Liu\textsuperscript{1}, Hao Li\textsuperscript{2,3}, Yang Yang\textsuperscript{1}, Fei Chen\textsuperscript{2,3}, Xueliang Zhang\textsuperscript{1,$^{\dagger}$}\thanks{$^{\dagger}$ Corresponding author. }
\end{tabular}
}

\address{
\textsuperscript{1}College of Computer Science, Inner Mongolia University, China\\
\textsuperscript{2}Department of Electrical and Electronic Engineering, \\Southern University of Science and Technology, Shenzhen, China\\
\textsuperscript{3}Key Laboratory of Urban Land Resources Monitoring and Simulation, \\Ministry of Natural Resources, China\\
\small \texttt{heshulin@mail.imu.edu.cn,cszxl@imu.edu.cn}
}

\begin{document}
\ninept
\setstretch{0.84} 
\maketitle
\begin{abstract}

Target speaker extraction (TSE) aims to isolate a specific voice from multiple mixed speakers relying on a registerd sample. Since voiceprint features usually vary greatly, current end-to-end neural networks require large model parameters which are computational intensive and impractical for real-time applications, espetially on resource-constrained platforms. In this paper, we address the TSE task using microphone array and introduce a novel three-stage solution that systematically decouples the process: First, a neural network is trained to estimate the direction of the target speaker. Second, with the direction determined, the Generalized Sidelobe Canceller (GSC) is used to extract the target speech. Third, an Inplace Convolutional Recurrent Neural Network (ICRN) acts as a denoising post-processor, refining the GSC output to yield the final separated speech. Our approach delivers superior performance while drastically reducing computational load, setting a new standard for efficient real-time target speaker extraction.
\end{abstract}
\section{Introduction}
\label{sec:intro}
Unlike traditional speech enhancement systems, target speaker extraction emerges as a highly practical solution to the well-known cocktail party problem \cite{vzmolikova2019speakerbeam}.
In this approach, the system is designed to exclusively extract the speech of a designated target speaker from a mixture of voices, using the reference speech of that target speaker as a guide.
This task holds a pivotal role in various applications, including voice recognition, multi-speaker transcription, and assistive hearing technologies \cite{wang2018voicefilter}.
It effectively amplifies the voice of the target speaker while concurrently suppressing interfering voices, thereby enhancing the clarity and intelligibility of the desired speech signal.
Inspired by human auditory attention, current deep learning speaker extraction methods \cite{mesgarani2012selective,kaya2017modelling,deng21c_interspeech,han2021attention,li2019dynamic} predominantly employ a two-subnet architecture comprising a speaker encoder and speech extractor \cite{xu2019time,delcroix2020improving,ju2023tea3,zhang23r_interspeech,lin23b_interspeech}. The speaker encoder models the voiceprint of the target speaker and subsequently guides the speech extractor to isolate the speech signal associated with that specific speaker.
However, the voiceprint extracted by the encoder and the mixture signal features occupy disparate high-dimensional spaces.
Bridging this gap requires extensive end-to-end neural networks, which are often complex and large \cite{chen2023mc,ju2023tea3}.

For the first time, Voicefilter \cite{wang2018voicefilter} uses the speaker's d-vector as an index to successfully extract the target speaker from the multi-talker speech, improving the recognition accuracy for different speakers.
To address the inherent difficulty associated with phase estimation in target speaker extraction, Xu et al. \cite{xu2020spex} integrate a time domain model into the extraction task, resulting in significant performance improvements.
Despite these advancements, the disparity between the target speaker's frequency domain voiceprint and the time domain mixed speech feature space remains a challenging issue.
Zmolikova et al. \cite{vzmolikova2019speakerbeam} propose joint training of speaker and separation networks to reduce the feature space disparity between voiceprint and mixed speech. 
Similarly, Ge et al. \cite{ge2020spex+} propose joint training in the time domain to reduce feature disparity.
He et al. \cite{he2020Speakerfilter} improve registration speech processing via multi-level feature extraction to better utilize registration information.
Additionally, Ge et al. \cite{ge2022spex} expand the scope of the target speaker extraction problem to encompass multi-microphone scenarios.
However, due to their model’s reliance on Bi-directional Long Short-Term Memory (BLSTM) networks and global Minimum Variance Distortionless Response (MVDR) beamforming, the resultant model parameters are substantial, posing challenges for real-time applications.

Unfortunately, the trend in literature is to develop increasingly large and complex models in pursuit of higher accuracy and precision, often at the expense of computational efficiency.
This focus on performance has overshadowed the need for models that can operate effectively under real-time, low-resource conditions \cite{10379131}, such as those required by edge computing devices.
To the best of our knowledge, there exists no prior work addressing the problem of target speaker extraction under real-time and resource-constrained conditions, such as those encountered in edge computing environments.

In this work, we introduce 3S-TSE: three-stage target speaker extraction technique tailored for multi-microphone environments.
Rather than extracting the target speaker's voice, the first stage is to estimate the target speaker's direction of arrival (DOA) by an encoder based on attentive recurrent network. 
This strategy shifts the paradigm from an approximation-based challenge to a classification task, effectively mitigating the intrinsic complexity.
In second stage, using the DOA determined in the first stage, we employ the Generalized Sidelobe Canceller (GSC) \cite{griffiths1982alternative} to preliminarily segregate the target speaker's voice from the mixed signals.
In order to further suppress the residual in the output of GSC, we propose a post-processing module based on Inplace Convolutional Recurrent Network (ICRN).
By adopting this decoupled strategy for multi-microphone target speaker extraction, we have successfully reduced the overall difficulty of the task.
On the Librispeech dataset, our approach remarkably achieved a 17.3\% improvement in short-time objective intelligibility (STOI) while maintaining a compact model with only 0.19M parameters.
 
The main contributions of this work are three-fold:
1) we introduce a Three-Stage approach for target speaker extraction, segregating the process into direction of arrival estimation, beamforming, and post-processing. This method not only has excellent performance but also has an ultra-compact model.
2) For traditional beamforming, we propose an efficient post-processing method based on inplace convolution. This module is notable for its exceptional performance, achieved with minimal computational load.
3) Our method achieves a significant objective improvement, registering a 17.3\% improvement in short-time objective intelligibility (STOI), while utilizing a minimal model size of just 0.19M parameters.

\section{Methodology}
\label{sec:PROPOSED 3S-TSE}

\subsection{Notations}
\label{ssec:Notations}
\begin{figure}[ht!]
	\centering
	\includegraphics[width=8.5cm]{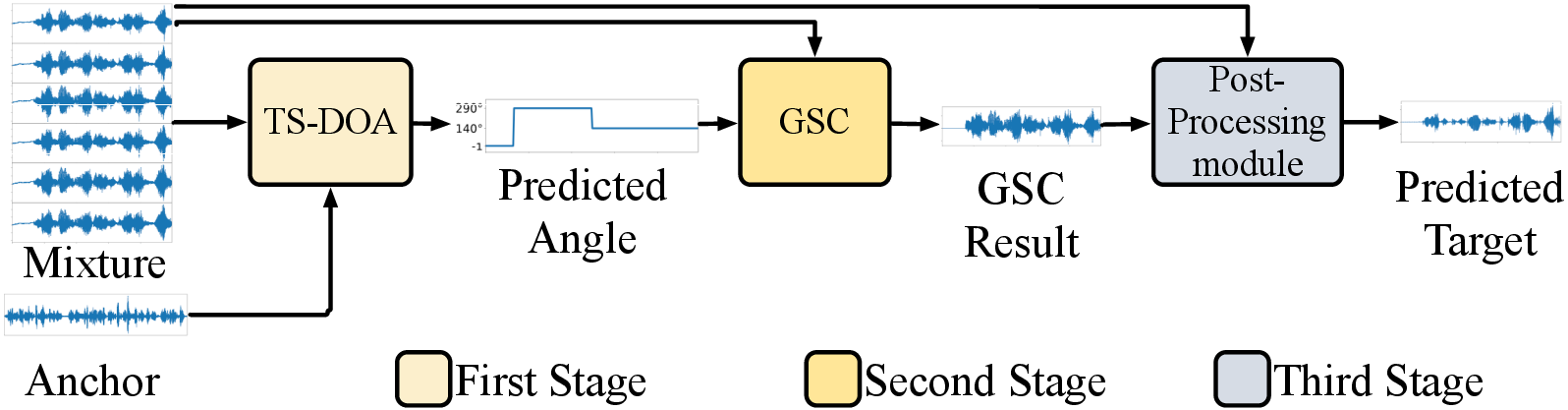}
	\caption{Overview of the proposed 3S-TSE system.}
	\label{fig:structure}
\end{figure}
As illustrated in Fig. \ref{fig:structure}, we present a comprehensive diagram of our proposed approach.
This section begins with an overview of our microphone array structure and subsequently provides detailed descriptions of its key components: the Target Speaker Direction of Arrival Estimation Module (TS-DOA), the Generalized Sidelobe Canceller (GSC) algorithm, and the Post-Processing Module.

\subsection{System overview}
\label{ssec:System overview}

\begin{figure}[ht!]
	\centering
	\includegraphics[width=8.5cm]{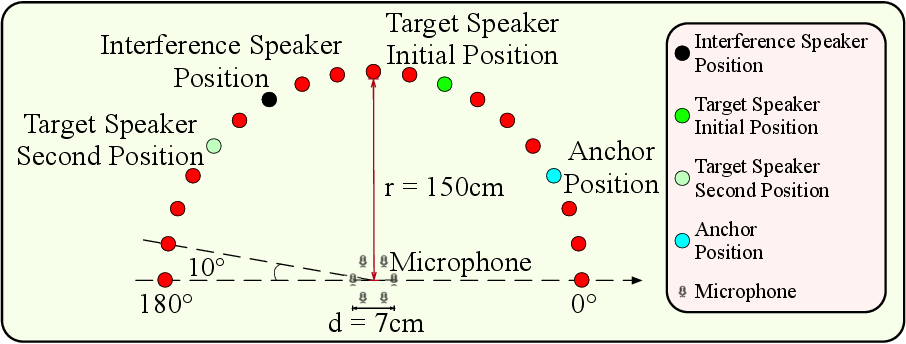}
	\caption{Structure of the six-channel microphone array. For brevity, the figure presents the source signal distribution from 0° to 180°. The unseen distribution from 180° to 360° is a mirror reflection of the shown pattern, ensuring symmetry.\vspace{-0.2cm}}
	\label{fig:array overview}
\end{figure}
As depicted in Fig. \ref{fig:array overview}, our proposed speech enhancement system comprises a circular array of six omnidirectional microphones, denoted as $M_1, \ldots, M_6$.
This array, with a radius of $3.5$ cm, is designed to amplify speech from the target speaker's direction and suppress interference from other directions.

We train the system using 36 source locations, spaced evenly from 0° to 360° at 10° intervals, with each location 1.5 m from the center of the microphone array.
For each training example, we randomly select four distinct locations from the 36 source locations.
These four locations represent the initial position, second position, interfering position, and anchor position for the target speaker.
To simulate the movement of the target speaker, we randomly shift the speaker from the initial position to the second position at a randomly chosen time point during each utterance simulation.

Our proposed framework encompasses three main stages, as depicted in the Fig. \ref{fig:structure}: the first stage is TS-DOA, followed by the second stage which utilizes the GSC algorithm, and the third stage which is the post-processing module. The TS-DOA determines the angle of the target speaker for each frame. With the target angle and the six-channel mix as inputs, the GSC algorithm generates an initial estimate of the target speech. This initial target speech estimate, along with the six-channel mix, is then forwarded to the post-processing module. Within this module, each TF-bin undergoes refinement, ultimately yielding enhanced speech.


\vspace{-0.3cm}
\subsection{First Stage: TS-DOA}
\label{ssec:TS-DOA}

\begin{figure}[ht!]
	\centering
	\includegraphics[width=8.5cm]{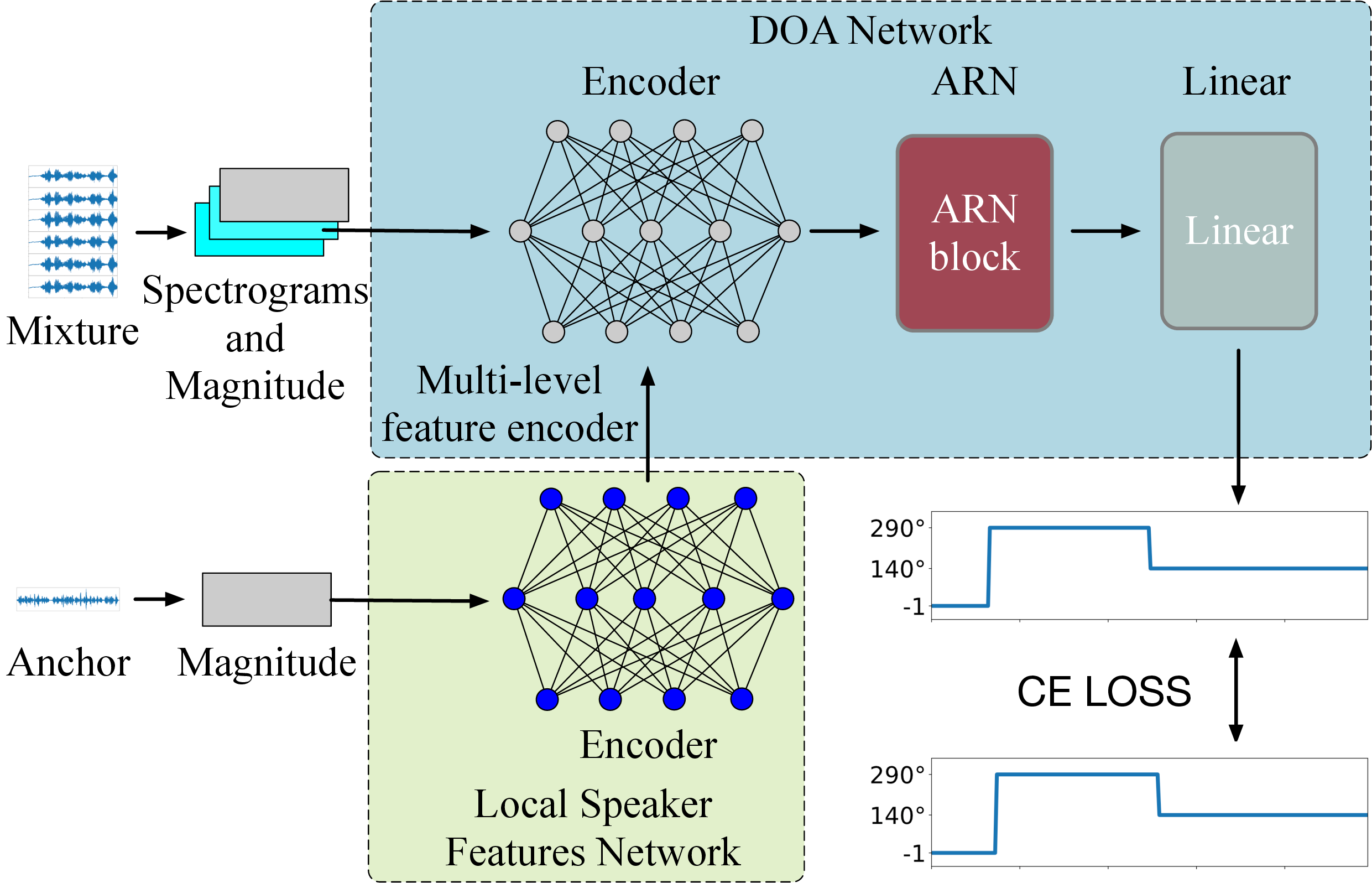}
	\caption{(First Stage) Block Diagram of TS-DOA Module. Note: an output angle value of -1 indicates the absence of the target speaker's voice in that frame.\vspace{-0.3cm}}
	\label{fig:ts-doa}
\end{figure}
The TS-DOA module in our framework consists of two major components: a local speaker feature network and a DOA network.
The DOA network utilizes an encoder-temporal architecture.
In this configuration, the magnitude and complex spectrograms of the mixed speech are used as inputs.
This encoder is composed of five convolutional layers, with the number of kernels in each layer increasing progressively.
To aggregate contextual information along the frequency axis, a stride of 2 is implemented in the convolutional layers along this dimension.
Following the encoding stage, an Attention Recurrent Network (ARN) \cite{pandey2022self} layer is introduced to model the temporal dependencies within the features.
A linear layer is employed as the output layer, responsible for producing the final estimated direction of the target speaker.
To enhance the model's generalization capability, we incorporate dropout regularization in both the ARN layer and the output layer. The dropout rate is set at 0.3.

The local speaker feature network has the same architecture as the encoder network in the DOA module.
First, the magnitude spectrum from the first channel of the anchor speech is encoded.
This encoded magnitude spectrum is combined with the output of each layer of the speaker encoder to form the local speaker features.
We compute the average across all frames for each feature.
The averaged local speaker features are concatenated, layer-by-layer, with the matching encoder layers in the DOA module for each time frame \cite{he2020Speakerfilter}, as depicted in Fig. \ref{fig:ts-doa}.
\vspace{-0.3cm}


\subsection{Second Stage: GSC}
\label{ssec:GSC}
Griffiths introduced the GSC, which adapts the linearly constrained minimum-variance proposed by Frost \cite{griffiths1982alternative} to an unconstrained scenario.
$[y_1(n), y_2(n), \ldots, y_M(n)]$ represents the speech inputs from the M microphones. The GSC \cite{gannot2001signal} structure primarily comprises two subsystems:
1. The upper branch houses the beamformer, characterized by a delay alignment structure and a weight set $W_c=\left[w_{c1}, w_{c2}, \ldots, w_{cM}\right]$. By manipulating $W_c$, one can control both the gain and beam width of this beamformer.
2. The lower branch contains the blocking matrix $B$, responsible for noise component estimation.
The outputs from these two subsystems undergo adaptive cancellation via an adaptive canceller paired with a coefficient matrix $W_a$, producing the enhanced speech.
Griffiths used the traditional least mean square (LMS) algorithm to derive the adaptive weights $W_a$. Here, $y_a(n)$ is the beamformer output, $y_z(n)$ denotes the noise in $y_a(n)$, and $y_o(n)$ represents the enhanced speech.

The output angle from the TS-DOA module is utilized as the target angle in GSC for computing the beamforming weight $W_a$.
The output of GSC is sent to the subsequent Post-Processing module as the preliminary target speaker extraction result.

\subsection{Third Stage: The Post-Processing module}
\label{ssec:The Post-Processing module}
To mitigate the limitations of GSC in handling nonlinear interference and reverberation, we implement a Post-Processing module.
This module takes a total of 7 channels of speech as input: the original six-channel mixture and the preliminary speech extraction result produced by the GSC algorithm.
The preliminary extracted speech from GSC maintains time-frequency correspondence with the original multi-channel mixture, ensuring aligned feature spaces.  
Importantly, since the GSC output is in the same feature domain as the original multi-channel input, no cross-domain mapping is required.

To process the time-frequency bin correspondence between the GSC output and microphone mixture, we implement inplace convolution \cite{liu2021inplace}, designed for efficient analysis of these alignments.
Unlike the classic CRN, which downsamples features along the frequency dimension, inplace convolution is designed with a stride set to one, thereby maintaining the original number of features.
This simple yet effective approach serves dual purposes: it retains key information for each target frequency bin and preserves the correlation of adjacent channels.
In our current task, inplace convolution processes each time-frequency point independently, avoiding the need for downsampling and thus preserving the characteristics of the original signal.
This preservation of signal characteristics is beneficial for the LSTM network, enabling it to effectively track the target speaker information across each frequency band.
Additionally, the inplace design minimizes parameters and computations, making it well-suited for this task's efficiency needs.

\begin{figure}[ht!]
	\centering
	\includegraphics[width=8.5cm]{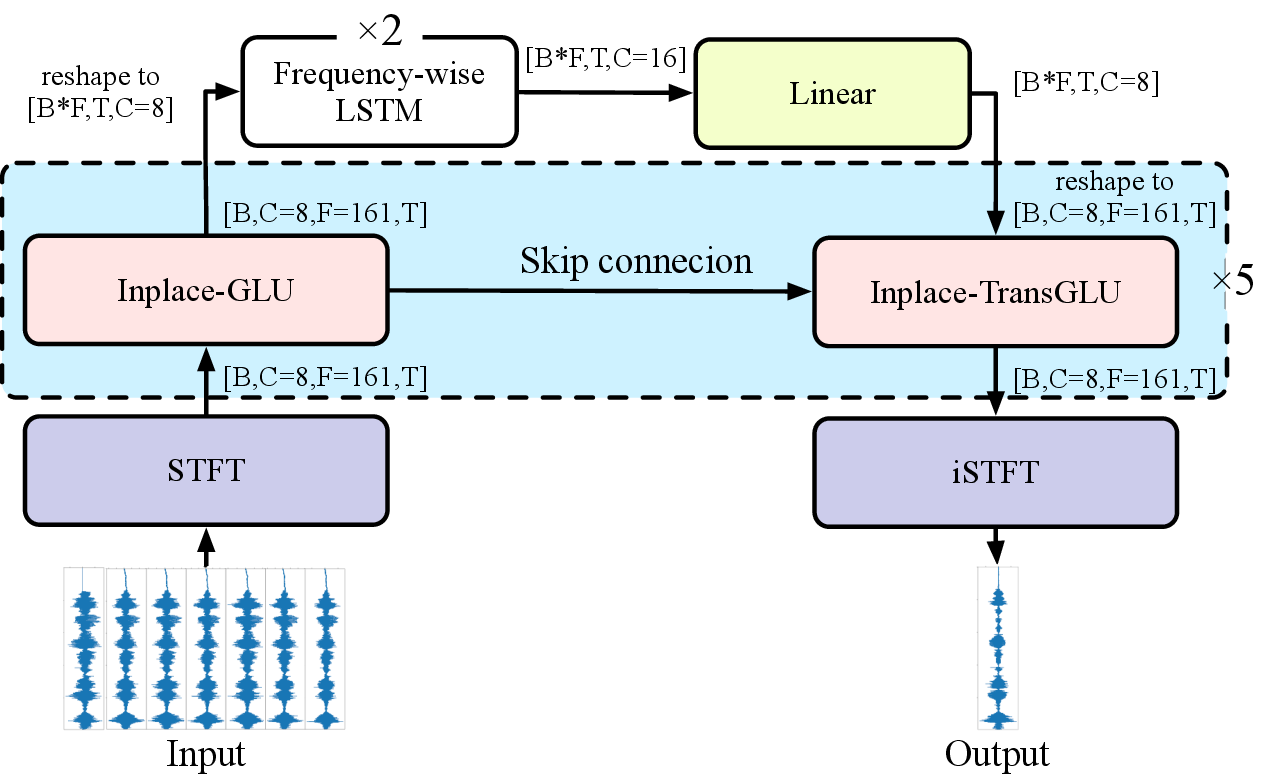}
	\caption{(Three Stage) Block Diagram of Inplace CRN Post-Processing Module.\vspace{-0.3cm}}
	\label{fig:pp}
\end{figure}

The Post-Processing module contains three main components: an encoder, a decoder, and a timing modeling unit, as depicted in Fig. \ref{fig:pp}.
The encoder and decoder each consist of 5 cascaded inplace convolution gated linear unit (GLU) layers, respectively.
Skip connections are employed to link the output of each inplace-GLU block directly to the input of the corresponding inplace-TransGLU block.
Sequentially following each inplace-GLU and inplace-TransGLU block are a batch normalization operation and an Exponential Linear Unit (ELU) activation function.
The decoder concludes with a linear layer, which is designed to project the learned features into a specified feature space.

For temporal modeling, we employ a two-layer Long Short-Term Memory (LSTM) network for each frequency component of the encoder's output, which we refer to as a frequency-wise LSTM.
This design choice means that each frequency is modeled independently, allowing for more fine-grained temporal dynamics to be captured.
To limit parameters and maintain compactness, the LSTM model is shared across frequencies. Unlike the Convolutional Recurrent Network (CRN) which processes full spectrum input, our frequency-wise LSTM enables focused subband modeling.
This subband approach focuses modeling on particular frequencies, enabling more precise and interpretable temporal representations.

\subsection{Loss function}
\label{ssec:LOSS FUNCTION}

For the TS-DOA task, we employ a cross-entropy loss \cite{shore1980axiomatic}, contrasting the true class labels with predicted probabilities derived from a softmax function. For the target speaker extraction task, the model is optimized using a combination of losses: the Scale-Invariant Signal-to-Noise Ratio (SI-SNR) loss \cite{luo2019conv}, which gauges signal quality by comparing estimated and original clean sources; a magnitude MSE loss for spectrum accuracy; and a Real and Imaginary (RI) MSE loss to assess phase differences \cite{ju2023tea3}. 

\section{Experiments}
\label{sec:Experiments}

\subsection{Dataset and setup}
\label{ssec:ataset}

For the training and test datasets, we generate six-channel mixtures by simulating a reverberant room environment and placing target and interfering speakers at different locations within the room.
This reverberant environment is modeled using a room with dimensions $(5 m \times 6 m \times 3 m)$, within which we generate a reverberant signal by convolving a speech signal with a six-channel room impulse response (RIR).
The RIRs are generated using the Image method, with a reverberation time (T60) randomly selected from $\{0.2, 0.3, 0.4, 0.5, 0.6, 0.7\}$ seconds.

We generate training and test mixtures using speech data from the LibriSpeech corpus. The training set consists of 20,000 two-speaker mixed utterances simulated as described in Section \ref{ssec:System overview}.
Specifically, we randomly select the voices of two different speakers to form each mixture.
Within each 2-talker mixed speech, we designate each speaker as the target speaker in turn, and the corresponding auxiliary reference speech is randomly selected from the original LibriSpeech corpus.
The Signal-to-Noise Ratio (SNR) between the target speech and the interfering speech is randomly sampled in a range from -5dB to 5dB, with a step size of 1dB.
For evaluation and comparison, we generate a separate set of 600 mixtures.
Additionally, we construct a test set comprising 1,000 mixtures, ensuring that the speech signals used in this set are distinct from those in the training set.
In all utterances, the duration of the shortest speech is 3 seconds, while the longest is 16 seconds.
All the utterances are sampled at a rate of 16 kHz.

For the short-time Fourier transform (STFT), we use a 20 ms Hanning window with a 10 ms shift.
With the 16 kHz sampling rate, a 320-point discrete Fourier transform (DFT) is applied to extract 161-dimensional complex spectra. 
The model is optimized using Adam with an initial learning rate of 0.001.
When the validation loss stops decreasing for two consecutive epochs, the learning rate is halved.
We use a batch size of 8 during training.

As we are performing online per-frame DOA estimation of the target speaker, silence segments are labeled without a direction. This results in 37 possible output targets per frame: 36 direction classes and a 1 no-direction silence class. Voice activity detection (VAD) is applied on the clean target speech to differentiate silent frames from voiced speech.

We use two metrics to evaluate DOA estimation performance: 1) Voice decision error (VDE) indicates the percentage of frames misclassified as either voice or silence. 2) Accuracy ratio (AR) \cite{lee2012noise} calculates the percentage of voiced frames where the estimated DOA is within $\pm 10^{\circ}$ of the ground truth angle.

To evaluate the performance of the target speaker extraction task, four objective metrics are employed: short-time objective intelligibility (STOI) \cite{STOI}, extended short-time objective intelligibility (ESTOI) \cite{jensen2016algorithm}, and perceptual evaluation of speech quality (PESQ) \cite{PESQ}.
Higher numbers indicate better performance for the AR, STOI, ESTOI, and PESQ metrics.
For the VDE, smaller numbers indicate better performance.

\begin{table*}[!htb] 
  \footnotesize
  \centering
  \caption{Comparative Analysis of Models Across Different Reverberation (T60) Levels.}
    \begin{tabular}{lcccccccccccc}
    \toprule
    T60   & \multicolumn{2}{c}{0.2} & \multicolumn{2}{c}{0.3} & \multicolumn{2}{c}{0.4} & \multicolumn{2}{c}{0.5} & \multicolumn{2}{c}{0.6} & \multicolumn{2}{c}{0.7} \\
    \cmidrule(lr){2-3}\cmidrule(lr){4-5}\cmidrule(lr){6-7}\cmidrule(lr){8-9}\cmidrule(lr){10-11}\cmidrule(lr){12-13}
    Metrics  & ESTOI & PESQ  & ESTOI & PESQ  & ESTOI & PESQ  & ESTOI & PESQ  & ESTOI & PESQ  & ESTOI & PESQ \\
    \midrule
    Mixture & 0.503  & 1.720  & 0.464  & 1.697  & 0.426  & 1.624  & 0.389  & 1.575  & 0.366  & 1.500  & 0.341  & 1.476  \\
    GT\_GSC & 0.632  & 2.122  & 0.562  & 1.979  & 0.511  & 1.858  & 0.464  & 1.739  & 0.447  & 1.690  & 0.417  & 1.621  \\
    TS-DOA+GSC & 0.550  & 1.706  & 0.492  & 1.614  & 0.449  & 1.544  & 0.406  & 1.477  & 0.392  & 1.411  & 0.392  & 1.411  \\
    TS-DOA(large)+GSC & 0.591  & 1.920  & 0.529  & 1.788  & 0.484  & 1.714  & 0.439  & 1.620  & 0.422  & 1.564  & 0.395  & 1.553  \\
    LGR   & 0.463  & 1.596  & 0.434  & 1.571  & 0.406  & 1.541  & 0.381  & 1.558  & 0.376  & 1.502  & 0.350  & 1.466  \\
    LGR(large) & 0.677  & 2.273  & 0.652  & 2.236  & 0.587  & 2.085  & 0.545  & 1.985  & 0.523  & 1.856  & 0.505  & 1.870  \\
    3S-TSE & 0.748  & 2.515  & 0.712  & 2.451  & 0.660  & 2.245  & 0.623  & 2.120  & 0.619  & 2.095  & 0.568  & 1.975  \\
    \bottomrule
    \end{tabular}%
  \label{tab:4}%
  \vspace{-0.3cm}
\end{table*}%
\begin{table}[htbp]
  \footnotesize
  \centering
  \caption{Performance Comparison of TS-DOA Models of Different Sizes on Direction of Arrival Estimation.}
    \begin{tabular}{lcccc}
    \toprule
    Metrics  & VDE   & AR    & Params(M) & MACs(G) \\
    \midrule
    TS-DOA & 0.1167  & 0.6973  & 0.037  & 0.038  \\
    \midrule
    TS-DOA(Large) & 0.0606  & 0.8358  & 0.444  & 0.730  \\
    \midrule
    TS-DOA(Huge) & 0.0595  & 0.8581  & 4.609  & 1.347  \\
    \bottomrule
    \end{tabular}%
  \label{tab:1}%
  \vspace{-0.3cm}
\end{table}%

\subsection{Experiment results}
\label{ssec:Experiment results}



Table \ref{tab:1} compares three versions of our proposed target speaker DOA model: TS-DOA, TS-DOA Large, and TS-DOA Huge. Model complexity is also compared by parameter count and Multiply-Accumulate operations (MACs).

As model size increases from TS-DOA to TS-DOA Huge, DOA estimation improves, with lower VDE and higher AR. 

The compact TS-DOA model achieves low complexity with 0.037M parameters and 0.038G MACs. Performance is moderate, with VDE of 0.1167 and AR of 0.6973. 
TS-DOA Large improves performance substantially, reducing VDE to 0.0606 and elevating AR to 0.8358. However, complexity rises to 0.444M parameters and 0.730G MACs.
Despite 4.609M parameters and 1.347G MACs, TS-DOA Huge only marginally enhances performance further. The scaling provides minimal gains.


\begin{table}[htbp]
  \footnotesize
  \centering
  \caption{Comparison of TS-DOA Models for Target Speaker DOA Estimation.}
   \setlength{\tabcolsep}{1mm}{ \begin{tabular}{lccccc}
    \toprule
    Metrics  & STOI  & ESTOI & PESQ  & Params(M) & MACs(G) \\
    \midrule
    Mixture & 0.620  & 0.419  & 1.605  & -     & - \\
    \midrule
    GT\_GSC & 0.713  & 0.510  & 1.846  & -     & - \\
    \midrule
    TS-DOA+GSC & 0.643  & 0.445  & 1.531  & 0.037  & 0.038  \\
    \midrule
    TS-DOA(Large)+GSC & 0.683  & 0.481  & 1.702  & 0.444  & 0.730  \\
    \bottomrule
    \end{tabular}%
    }
    \vspace{-0.2cm}
  \label{tab:2}%
\end{table}%

In Table \ref{tab:2}, we compare target speaker DOA approaches using the Generalized Sidelobe Canceller (GSC) for extraction. Models include our proposed TS-DOA+GSC and TS-DOA Large+GSC, along with end-to-end baselines LGR and LGR Large. 

Both TS-DOA+GSC and LGR show suboptimal performance when operating with smaller parameters and computational costs. However, TS-DOA+GSC manages to achieve its results with notably lower parameter count and computational burden compared to the end-to-end LGR model \cite{he2022local}. This is significant, highlighting the efficiency of the TS-DOA+GSC approach under constrained resources.
For example, as shown in Table 1, TS-DOA+GSC attains a STOI of 0.643, ESTOI of 0.445, and PESQ of 1.531 using only 0.037M parameters and 0.038G MACs. 





\begin{table}[htbp]
  \footnotesize
  \centering
  \caption{Comparative Analysis of DOA Estimation Models Enhanced with Post-Processing Modules.}
     \setlength{\tabcolsep}{1mm}{\begin{tabular}{lccccc}
    \toprule
    Metrics  & STOI  & ESTOI & PESQ  & Params(M) & MACs(G) \\
    \midrule
    Mixture & 0.620  & 0.419  & 1.605  & -     & - \\
    \midrule
    LGR(Huge) & 0.753  & 0.586  & 2.063  & 4.948  & 2.022  \\
    \midrule
    TS-DOA GSC+ & \multirow{2}{*}{0.705} & \multirow{2}{*}{0.542} & \multirow{2}{*}{1.884} & \multirow{2}{*}{0.808} & \multirow{2}{*}{1.046} \\
    CRN &       &       &       &       &  \\
    \midrule
    3S-TSE & 0.793  & 0.659  & 2.246  & 0.190  & 1.000  \\
    \bottomrule
    \end{tabular}%
    \vspace{-0.2cm}}
  \label{tab:3}%
\end{table}%
In table \ref{tab:3}, we can see that the proposed 3S-TSE method achieves strong performance on the speech quality metrics stoi, estoi, and pesq, outperforming both the baseline Mixture and the large LGR end-to-end model. Notably, 3S-TSE attains this with orders of magnitude fewer parameters and computational cost than LGR, demonstrating the efficiency of the modular approach. Furthermore, in the experiment labeled as TS-DOA GSC+CRN, we replaced the post-processing module in 3S-TSE with CRN. The results indicated that CRN is not well-suited for refining outputs on a per-TF-bin basis. Overall, these results validate the effectiveness of the proposed pipeline, leveraging the strength of GSC for distortionless preliminary extraction coupled with an efficient post-processing module for refinement. The modular design enables improved performance to end-to-end approaches with reduced computational requirements.

Table \ref{tab:4} highlights the performance of models under various reverberation conditions, as characterized by the reverberation time T60. A key finding is that the TS-DOA GSC+post-processing model outperforms other models, particularly in highly reverberant environments with high T60 values. This superior performance, coupled with the low computational complexity of the TS-DOA GSC model, makes it an attractive option when both high performance and computational efficiency are critical. The consistent improvements from adding the post-processing module emphasize its value in refining the audio output.

\section{CONCLUSIONS}
\label{sec:CONCLUSIONS}

We propose a novel three-stage target speaker extraction method called 3S-TSE for multi-microphone scenarios. Firstly, we employ an encoder-attentive recurrent network (encoder-ARN) to learn the spatial angle at which the target speaker is situated. After estimating the direction of arrival (DOA), we employ the Generalized Sidelobe Canceller (GSC) algorithm to further isolate the speaker's voice from the mixed audio signal. Then, a Post-Processing module, comprised of an Inplace Convolutional Recurrent Neural Network (InplaceCRN), is applied to refine and output the final separated audio. 3S-TSE achieves a 17.3\% enhancement in short-time objective intelligibility (STOI) using just 0.19M parameters.
\\
\textbf{Acknowledgments}: This research was supported by the China National Nature Science Foundation (No. 61876214).
This work was also supported by the Open Fund (KF-2022-07-009) of Key Laboratory of Urban Land Resources Monitoring and Simulation, Ministry of Natural Resources, China.
\newpage
\bibliographystyle{IEEEbib}
\bibliography{strings,refs}

\end{document}